\documentclass[aps,twocolumn,showpacs,floats]{revtex4}
\usepackage{graphicx}
\usepackage{color}

\begin{document}

\title{Three-dimensional topological insulators in the octahedron-decorated cubic lattice }
\author{Jing-Min Hou}
\email{jmhou@seu.edu.cn}
\author{Wen-Xin Zhang}
\author{Guo-Xiang Wang}

\affiliation{ Department of Physics, Southeast University, Nanjing,
211189, China }
\date{\today}

\begin{abstract}
We investigate a tight-binding model of the octahedron-decorated
cubic lattice with spin-orbit coupling. We calculate the band
structure of the lattice and evaluate the $Z_2$ topological indices.
According to the $Z_2$ topological indices and the band structure,
we present the phase diagrams of the lattice with different filling
fractions. We find that the $(1;111)$ and $(1;000)$ strong
topological insulators occur in some range of parameters at $1/6,
1/2$ and $2/3$ filling fractions. Additionally, the $(0;111)$ weak
topological insulator is found at $1/6$ and $2/3$ filing fractions.
We analyze and discuss the characteristics of these topological
insulators  and their surfaces states.

\pacs{73.43.-f, 71.10.Fd, 73.20.-r, 72.25.-b}
\end{abstract}
\maketitle

\section{Introduction}
Usually, different phases of matter can be classified using Landau's
approach according to their underling symmetries\cite{Landau}. In
1980s, the discovery of the quantum Hall effect changed physicists'
viewpoint on the classification of matter\cite{Klitzing}. The
quantum Hall states can be classified by a topological invariant,
now named the TKNN number\cite{Thouless} (equivalent to the first
Chern number), which is directly connected to the quantized Hall
conductivity, but they have the same symmetry.   Since the Hall
conductivity is odd under time reversal, the topological non-trivial
quantum Hall states can only occur when time reversal symmetry is
broken, which is performed by a magnetic field. In 1988, Haldane
also proposed a time reversal symmetry broken toy model without a
magnetic field to realize quantum Hall states\cite{Haldane}. All the
quantum Hall states have a gapped band structure in bulk and chiral
gapless edge states that are topologically protected.

Recently, the promising prospect of spintronics in technology
stimulates  physicists to generate spin current. Quantum spin Hall
effect was proposed to create spin current\cite{Bernevig,Kane}. The
quantum spin Hall states are non-trivial topological phases with
time reversal symmetry, which have a bulk gap and topologically
protected gapless
 helical edge states. For the above reason,
the quantum spin Hall states also called topological insulators.
Two-dimensional   topological insulators are characterized by a
$Z_2$ topological index $\nu=0,1$\cite{Kane2}. For a non-trivial
topological insulator the topological index has a value $\nu=1$
while $\nu=0$ for a trivial band insulator.  Therefore, a
topological insulator always has a metallic boundary when placed
next to a vacuum or an ordinary band insulator because topological
invariants cannot change as long as a material remains insulating.
The remarkable metallic boundaries of topological insulators may
result in new spintronic or magnetoelectric devices and  a new
architecture for topological quantum bits. In quantum spin Hall
phases, the spin-orbit coupling plays the role of the spin-dependent
effective magnetic field. The first real material, a HgTe quantum
well, supporting two-dimensional topological insulators was
predicted by Bernevig, {\it et al.}\cite{Bernevig2} and
experimentally conformed by K\"onig {\it et al.}\cite{Konig}.

\begin{figure}[ht]
\includegraphics[width=0.45\columnwidth]{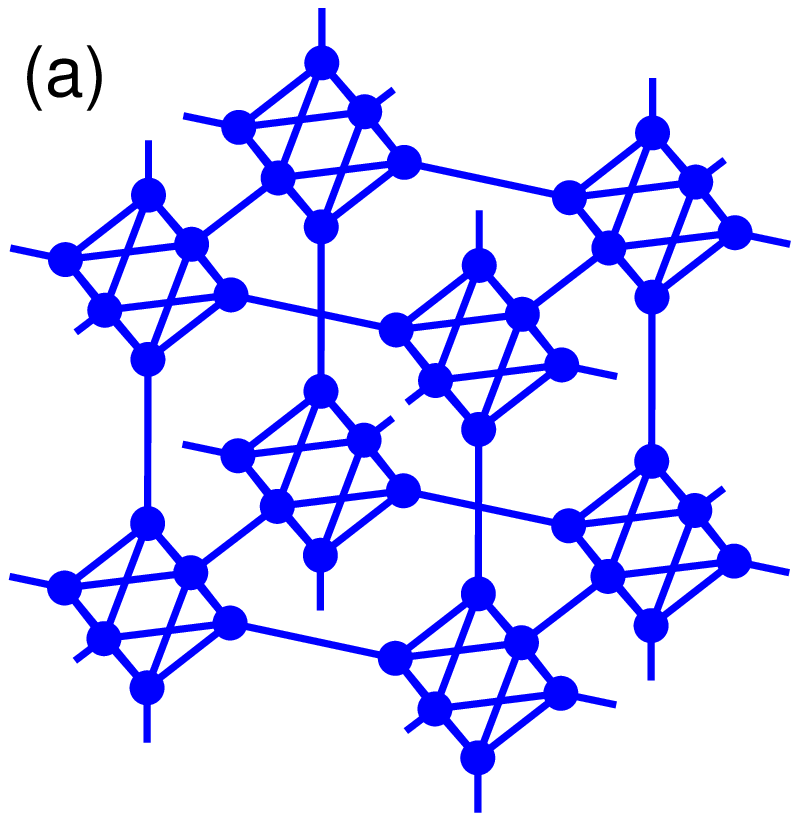}
\includegraphics[width=0.45\columnwidth]{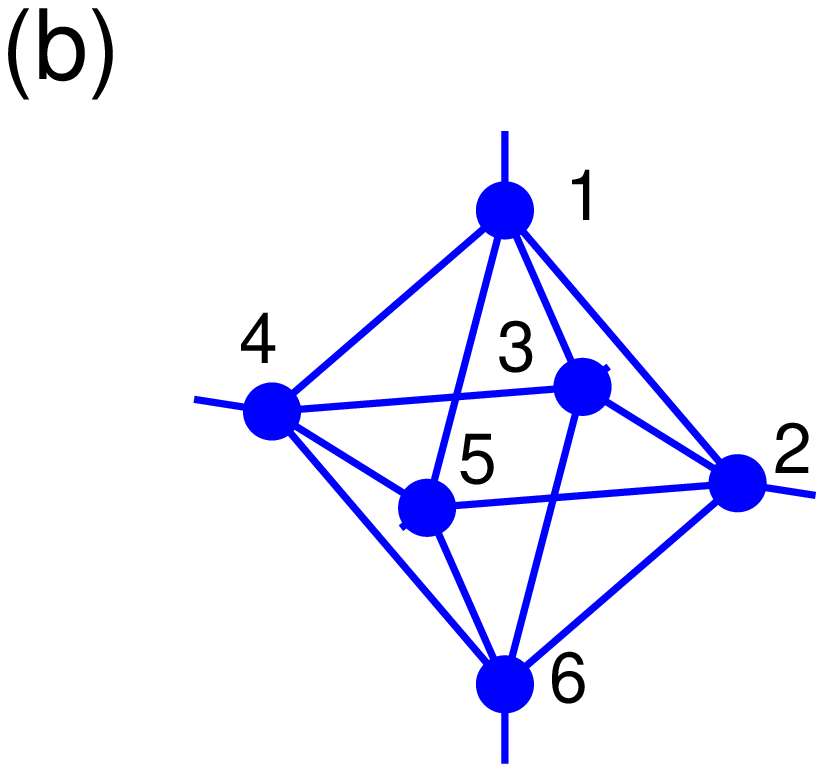}
\includegraphics[width=0.45\columnwidth]{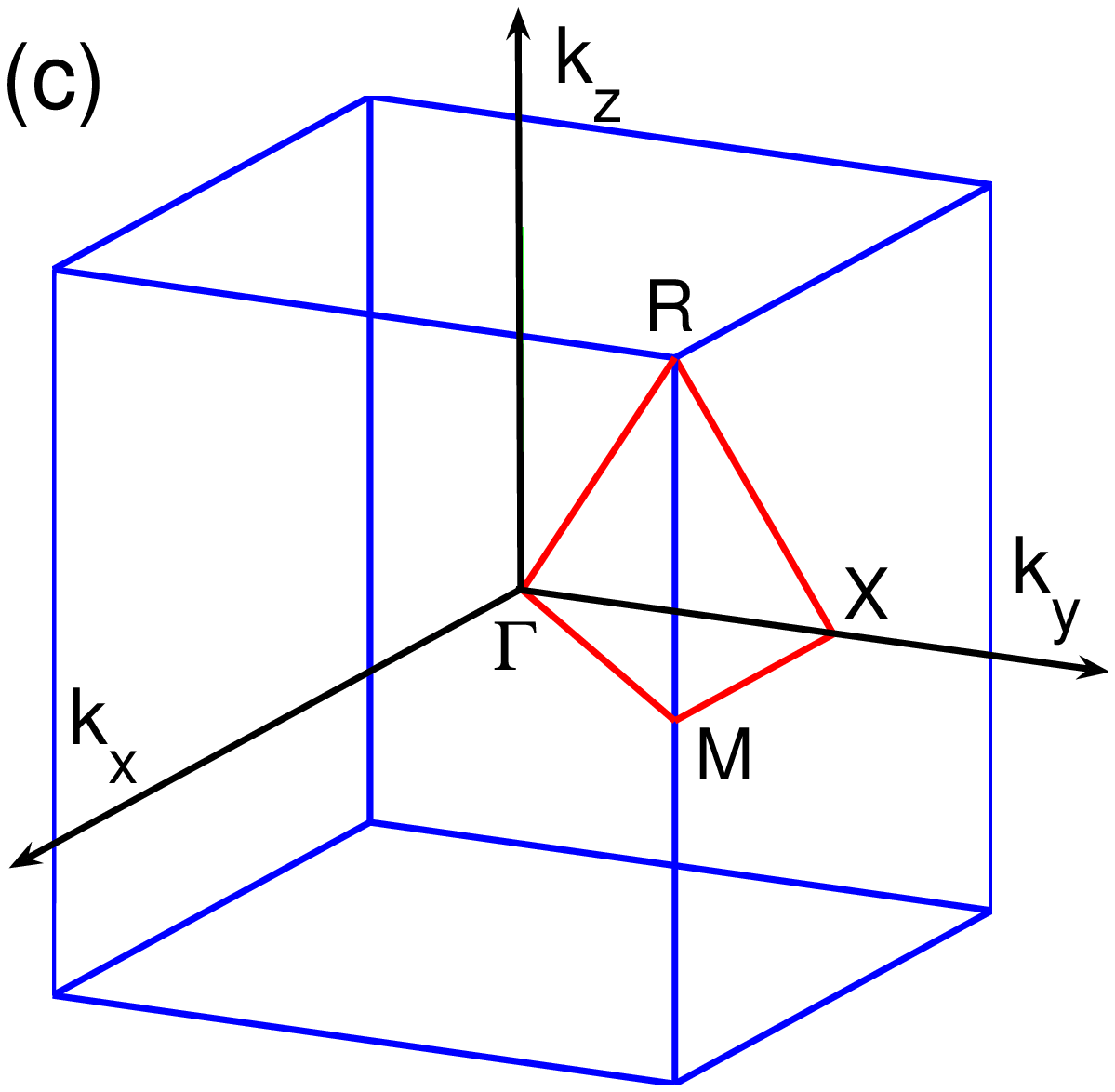}
\includegraphics[width=0.45\columnwidth]{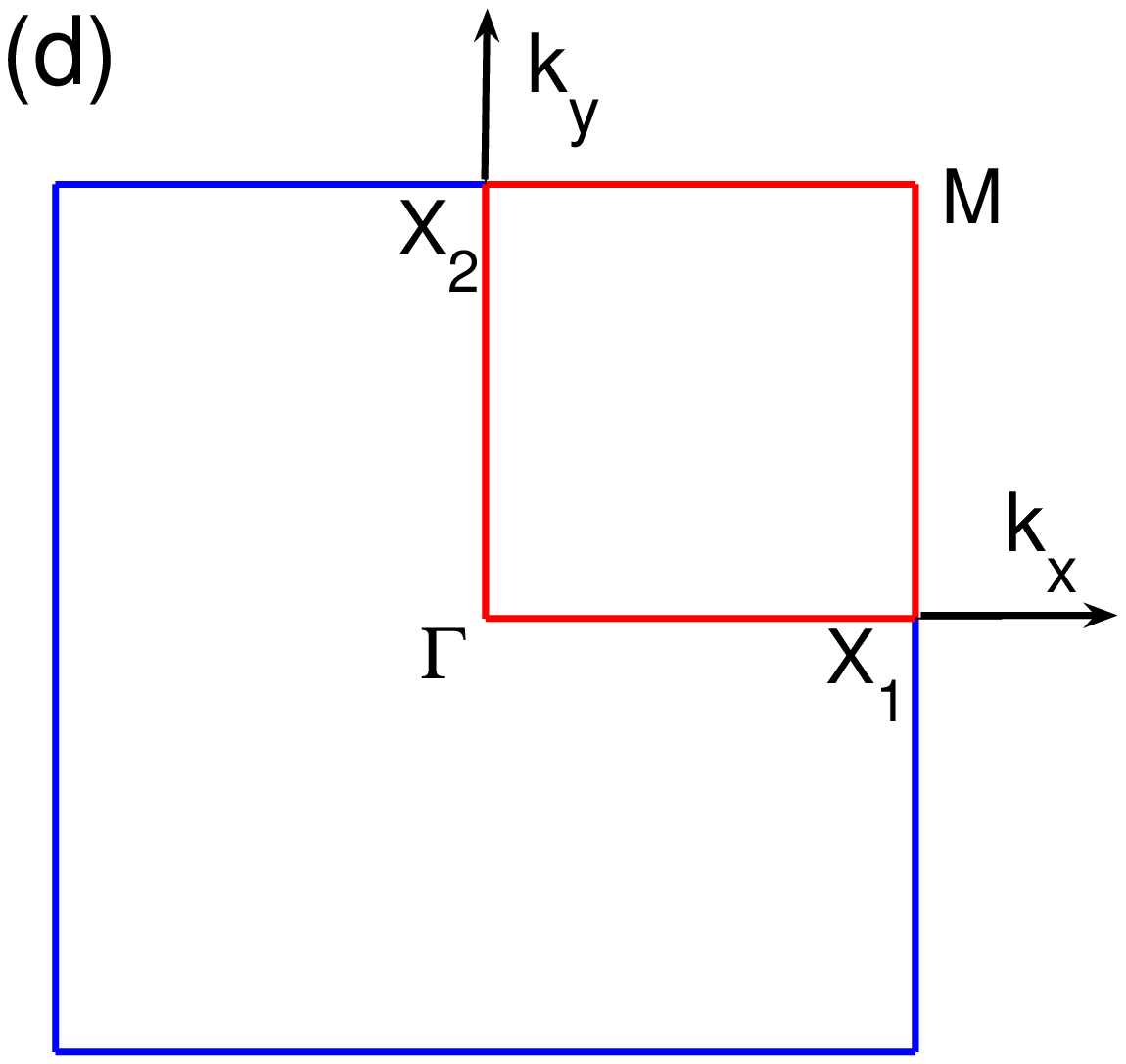}
\caption{(Color online). (a) The octahedron-decorated cubic lattice
which can be obtained by replacing every lattice site of a cubic
lattice with an octahedral cluster as shown in (b). (c) The
three-dimensional Brillouin zone and high symmetry points. (d) The
two-dimensional Brillouin zone of a slab with two $001$ surfaces.}
 \label{fig1}
\end{figure}

Soon after the quantum spin Hall insulator was discovered,
time-reversal invariant topological insulators were generalized to
three dimensions\cite{Fu,Moore,Roy}.
 Three-dimensional time-reversal
invariant band insulators are classified according to four $Z_2$
topological indices $(\nu_0;\nu_1\nu_2\nu_3)$ with
$\nu_i=0,1$\cite{Fu}. In three dimensions, the time-reversal
invariant band  insulators can be classified into 16 phases
according to the four $Z_2$ topological indices.  A band insulator
with $\nu_0=1$ is called a strong topological insulator(STI), a band
insulator with $\nu_0=0$ and at least one non-zero $\nu_i (i=1,2,3)$
is called a weak topological insulator(WTI), while an ordinary
trivial band insulator  has an index $(0;000)$. For an STI phase,
the surface states have an odd number of Dirac points, which are
topologically protected and for a WTI or trivial band insulator
phase, the surface states have an even number of Dirac points. Fu
and Kane firstly predicted that Bi$_{1-x}$ Sb$_x$ supports a
three-dimensional topological insulator\cite{Fu2}, which was
conformed experimentally by Hsieh, {\it et al}. in 2008\cite{Hsieh}.
Later, Bi$_2$Se$_3$ was discovered to be a three-dimensional
insulator experimentally as a second generation material\cite{Xia},
which also was  supported by theoretical
calculations\cite{Xia,H.Zhang}. Additionally, reference
\cite{H.Zhang} also predicted that Bi$_2$Te$_3$ and Sb$_2$Te$_3$ are
second generation materials supporting three-dimensional topological
insulators. The later  experimental studies  on
Bi$_2$Te$_3$\cite{Chen,Hsieh2,Hsieh3} and Sb$_2$Te$_3$\cite{Hsieh3}
identified their topological band structures.

To help experimental physicists find more topological insulator
materials, theoretical physicists have investigated several models
that support non-trivial topological insulators. Theoretical studies
have demonstrated that, within the tight-binding approximation and
with the spin-orbit coupling, the honeycomb\cite{Kane},
kagome\cite{Guo}, checkerboard\cite{Sun}, decorated
honeycomb\cite{Ruegg}, Lieb\cite{Weeks}, and
square-octagon\cite{Kargarian} lattices support two-dimensional
topological insulators and the diamond\cite{Fu},
pyrochlore\cite{Guo2}, and perovskite\cite{Weeks} lattices support
 three-dimensional topological insulators.

In this paper, we shall show that a new lattice, the
octahedron-decorated cubic lattice as shown in Fig.1 (a), supports
three-dimensional topological insulators with the spin-orbit
coupling existing. This lattice can be regarded as a
three-dimensional generalization of the square-octagon
lattice\cite{Ruegg}. We find that  this model supports STI and WTI
phases for $1/6$ and $2/3$ filling and STI phases for $1/2$ filling
as  well as ordinary band insulator and metal phases.

\section{Model}

We consider the octahedron-decorated cubic lattice as shown in
Fig.\ref{fig1} (a), which can be obtained by replacing every lattice
site of a cubic lattice with an octahedral cluster as shown in
Fig.\ref{fig1}(b). This lattice has a unit cell with six different
lattice sites as denoted in Fig.\ref{fig1}(b) so that it contains
six sublattices. Here, we assume that the distance between the
centers of two nearest-neighbor octahedral clusters is $a$, which is
the same with the lattice constant of all sublattices,  the distance
of every lattice site of an octahedral cluster  from its center is
$a/4$, and the distance of two nearest-neighbor lattice sites in
different octahedral clusters is $a/2$. With the tight-binding
approximation, we can write the second quantized Hamiltonian of the
lattice as follows,
\begin{eqnarray}
H_0=-t\sum_{\langle i,j\rangle , \sigma} c_{i\sigma}^\dag
c_{j\sigma}-t_1\sum_{[i,j], \sigma} c_{i\sigma}^\dag c_{j\sigma}
\label{H0}
\end{eqnarray}
where $c_{i\sigma}$ is  the annihilation operator destructing
  an electron with spin $\sigma$ on the site ${\bf r}_i$ of
the octahedron-decorated cubic lattice, $\langle i,j\rangle$
represents nearest-neighbor hopping in the same octahedral cluster
with amplitude $t$ and $[i,j]$ denotes nearest-neighbor hopping
between two different octahedral clusters with amplitude $t_1$.

\begin{figure}[ht]
\includegraphics[width=0.49\columnwidth]{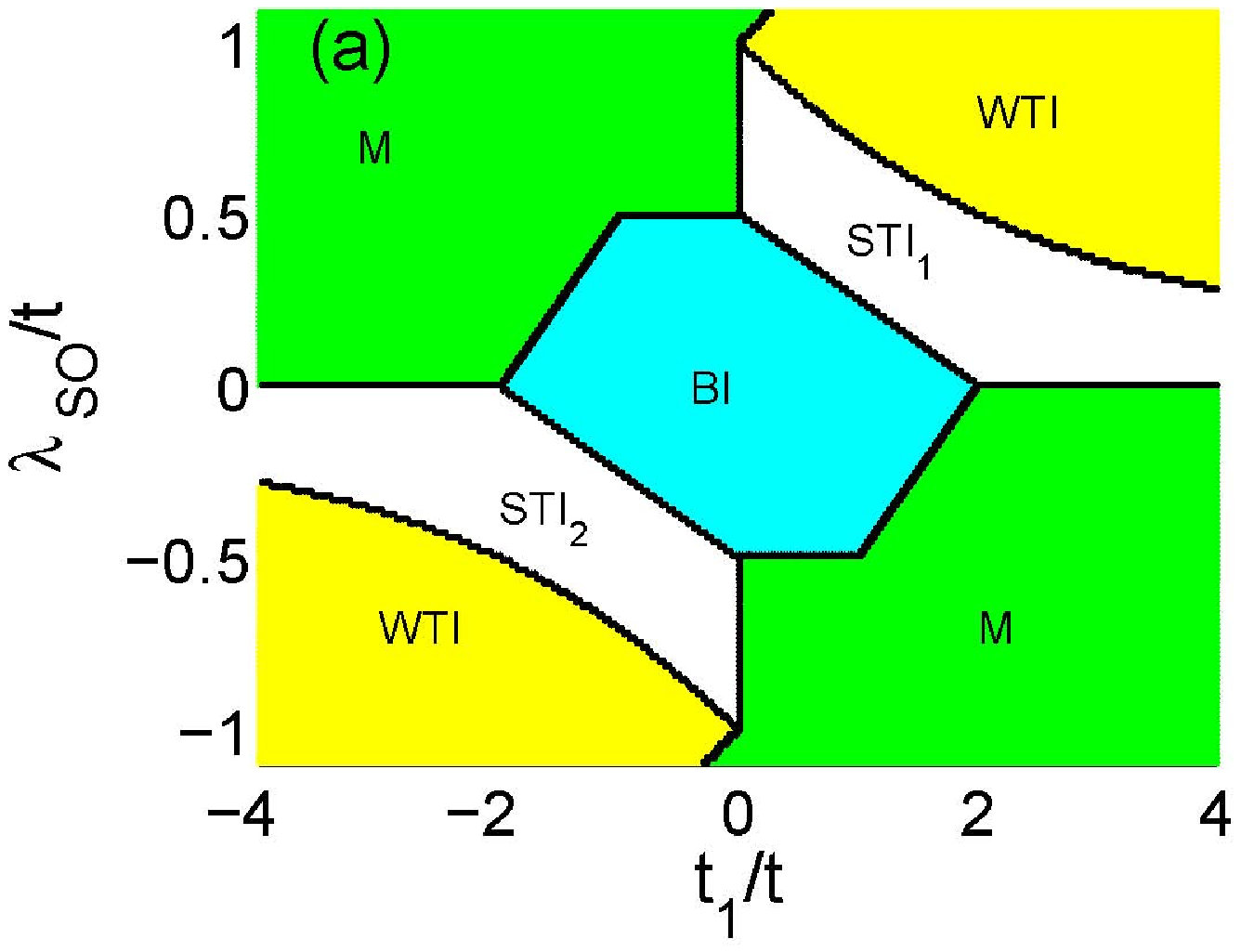}
\includegraphics[width=0.49\columnwidth]{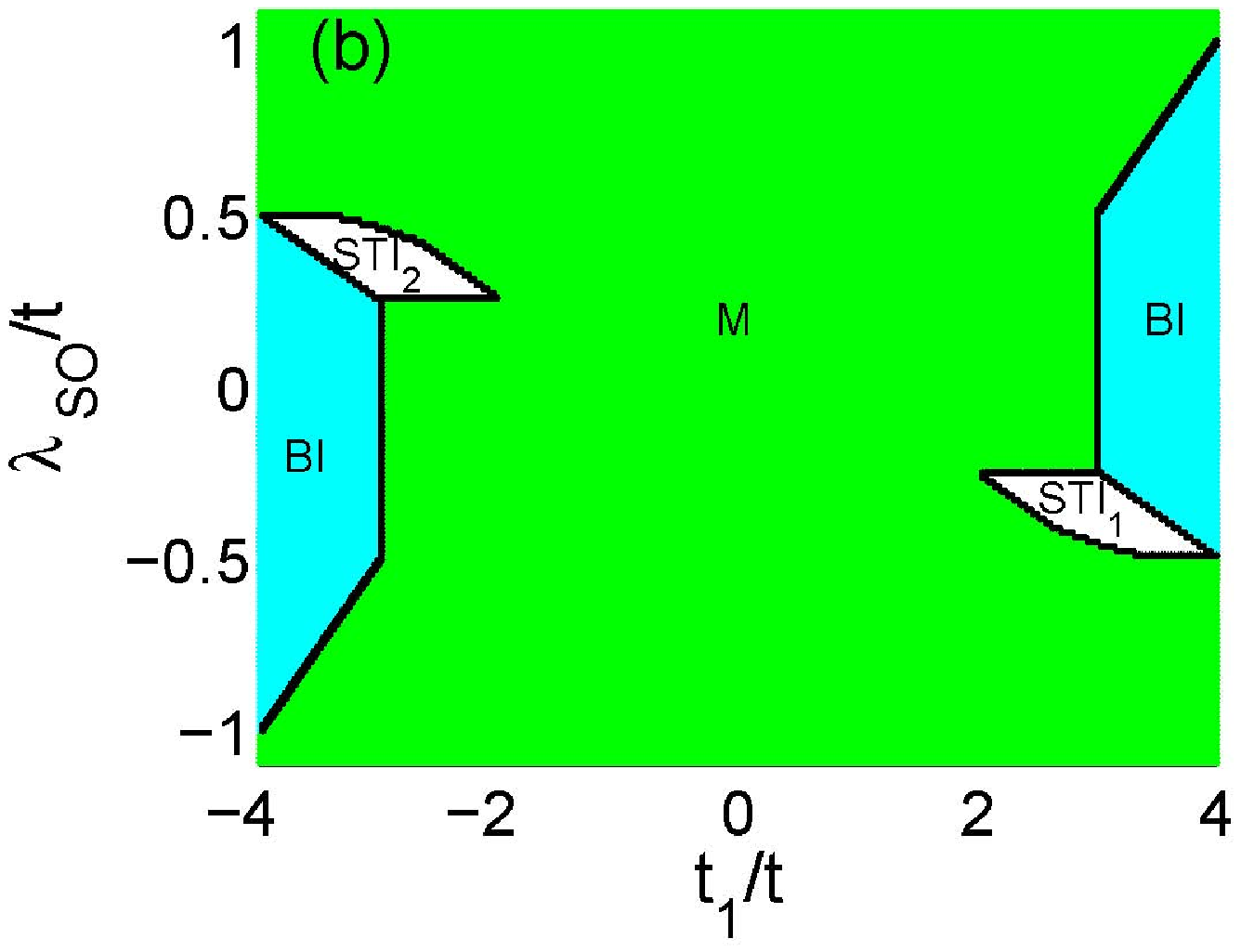}
\includegraphics[width=0.49\columnwidth]{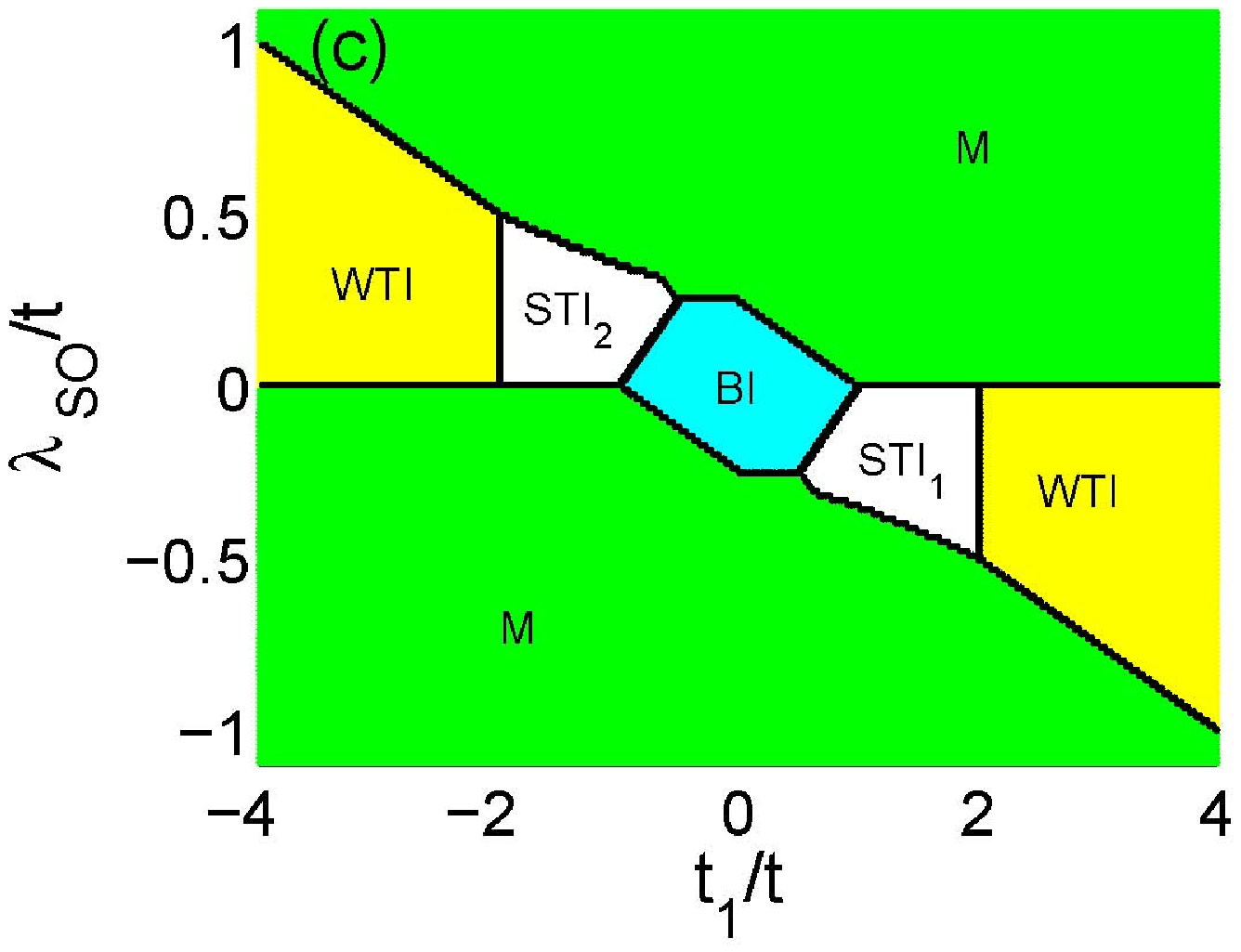}
\caption{(Color online). Phase diagrams of the octahedron-decorated
cubic lattice for (a) $1/6$ filling, (b) $1/2$ filling, and (c)
$2/3$ filling. Here,  BI  denotes a trivial band insulator; STI$_1$
and STI$_2$ denote $(1;111)$ and $(1;000)$ strong topological
insulators, respectively; WTI denotes a $(0;111)$ weak topological
insulator; and M denotes a metal   phase. } \label{fig2}
\end{figure}

In  momentum space, the Hamiltonian (\ref{H0}) can be represented by
$H_0=\sum_{{\bf k}\sigma}\Psi^\dag_{{\bf k}\sigma}{\cal
H}^{(0)}_{\bf k}\Psi_{{\bf k}\sigma}$ with $\Psi_{{\bf
k}\sigma}=(c_{1{\bf k}\sigma}, c_{2{\bf k}\sigma}, c_{3{\bf
k}\sigma}, c_{4{\bf k}\sigma},c_{5{\bf k}\sigma}, c_{6{\bf
k}\sigma})^T$,  which are ordered according to the sequence denoted
in Fig.\ref{fig1}(b). Here, ${\cal H}^{(0)}_{\bf k}$ takes  the
following form,
\begin{eqnarray}
&&{\cal H}_{\bf k}^{0}=\nonumber\\
&&-\left(\matrix{0&t&t&t&t&t_1e^{ik_z}\cr t&0&t&t_1e^{ik_x}&t&t\cr
t&t&0&t&t_1e^{ik_y}&t\cr t&t_1e^{-ik_x}&t&0&t&t\cr
t&t&t_1e^{-ik_y}&t&0&t\cr t_1e^{-ik_z}&t&t&t&t&0}\right)\nonumber\\
\label{Hk0}
\end{eqnarray}
Since $H_0$ is spin-decoupling, ${\cal H}_{\bf k}^{0}$ is
spin-independent, i.e. it is the same for both spin-up and spin-down
electrons. Fig.\ref{fig1}(c) shows the first Brillouin zone of the
octahedron-decorated cubic lattice. The spectrum of Eq.(\ref{Hk0})
with $t_1=t$ is calculated and shown in Fig.\ref{fig3}(d). The
spectrum contains six bands which come from the six sites in every
unit cell. A gap exists between the first and second bands.  The
second, third and fourth bands touch together at points $\Gamma, R$
and $M$. The third, fourth and fifth band touch at point $X$, near
which a Dirac cone occurs. Five bands including the second, third,
fourth, fifth and sixth bands meet at point $\Gamma$.  Along the
$\Gamma\rightarrow R$ line in momentum space, the second and third
bands are degenerate and the fifth and sixth bands are degenerate.

Now, in order to find non-trivial topological insulators in the
octahedron-decorated cubic lattice, we proceed to introduce the
spin-orbit interactions between next-nearest-neighbor sites as
follows,
\begin{eqnarray*}
H_{\rm SO}=i\frac{8\lambda_{\rm SO}}{a^2}\sum_{\langle\langle
i,j\rangle\rangle\alpha\beta}({\bf d}_{ij}^1\times{\bf
d}_{ij}^2)\cdot
\boldmath{\mbox{$\sigma$}}_{\alpha\beta}c_{i\alpha}^\dag c_{j\beta},
\label{HSO}
\end{eqnarray*}
where $\langle\langle i,j\rangle\rangle$ represents two
next-nearest-neighbor sites $i, j$, and $\lambda_{\rm SO}$ is the
amplitude of spin-orbit coupling of the two next-nearest-neighbor
sites.  ${\boldmath{\mbox{$\sigma$}}}=(\sigma_x, \sigma_y,\sigma_z)$
is the vector of Pauli spin matrices.  ${\bf d}_{ij}^{1,2}$ are the
two nearest neighbor bond vectors traversed between sites $i$ and
$j$ with $8|{\bf d}_{ij}^1\times{\bf d}_{ij}^2|/a^2=1$. In  momentum
space, the Hamiltonian for spin-orbit coupling (\ref{HSO}) can be
expressed as $H_{\rm SO}=\sum_{\bf k}\Psi_{\bf k}^\dag{\cal H}_{\bf
k}^{\rm SO}\Psi_{\bf k}$ with $\Psi_{\bf k}=(c_{1{\bf k}\uparrow},
c_{2{\bf k}\uparrow},$ $ c_{3{\bf k}\uparrow}, c_{4{\bf k}\uparrow},
c_{5{\bf k}\uparrow}, c_{6{\bf k}\uparrow}, c_{1{\bf k}\downarrow},
c_{2{\bf  k}\downarrow},  c_{3{\bf k}\downarrow}, c_{4{\bf
k}\downarrow},  c_{5{\bf k}\downarrow},  c_{6{\bf k}\downarrow})^T$.
Since ${\cal H}_{\bf k}^{\rm SO}$ does not decouple for the two spin
projections, it is a $12\times 12$ matrix. In momentum space, the
total single particle Hamiltonian is ${\cal H}_{\bf k}={\cal H}_{\bf
k}^0+{\cal H}_{\bf k}^{\rm SO}$. The bands and eigenstates can be
obtained by exactly diagonalizing ${\cal H}_{\bf k}$.

\section{Three-dimensional topological insulators  }

\begin{figure}[ht]
\includegraphics[width=1.1\columnwidth]{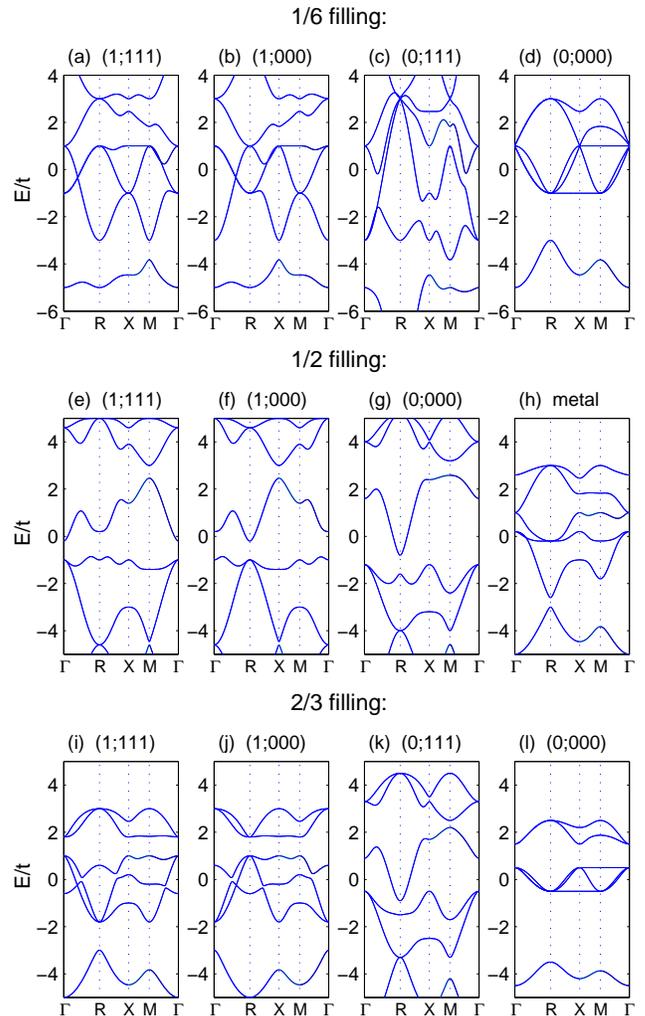}
\caption{(Color online). Band structures of the octahedron-decorated
cubic lattice for various parameters $t_1$ and $\lambda_{\rm SO}$.
Here, the horizontal axis represents the wave vectors along the path
in the first Brillouin zone indicated by the red lines in
Fig.\ref{fig1}(c). (a) $t_1=t, \lambda_{\rm SO}=0.5t$, (b) $t_1=-t,
\lambda_{\rm SO}=-0.5t$, (c) $t_1=t, \lambda_{\rm SO}=t$, (d)
$t_1=t, \lambda_{\rm SO}=0$, (e) $t_1=3t, \lambda_{\rm SO}=-0.4t$,
(f) $t_1=-3t, \lambda_{\rm SO}=0.4t$, (g) $t_1=3.2t, \lambda_{\rm
SO}=-0.2t$, (h) $t_1=t, \lambda_{\rm SO}=0.2t$, (i) $t_1=t,
\lambda_{\rm SO}=-0.2t$, (j) $t_1=-t, \lambda_{\rm SO}=0.2t$, (k)
$t_1=2.5t, \lambda_{\rm SO}=-0.2t$, and (l) $t_1=0.5t, \lambda_{\rm
SO}=0$. }\label{fig3}
\end{figure}

The classification of three-dimensional topological insulators is
presented in Ref.\cite{Fu}. For three-dimensional lattices there
eight distinct time reversal invariant momenta (TRIM), which can  be
expressed in terms of primitive reciprocal lattice vectors as
$\Gamma_{i=(n_1,n_2,n_3)}=(n_1{\bf b}_1+n_2{\bf b}_2+n_3{\bf
b}_3)/2$ with $n_j=0, 1$.  Three-dimensional topological insulators
can be distinguished by four $Z_2$ topological invariants $(\nu_0;
\nu_1 \nu_2 \nu_3)$, which are defined as
$(-1)^{\nu_0}=\prod_{n_j=0,1}\delta_{n_1n_2n_3}$ and
$(-1)^{\nu_{i=1,2,3}}=\prod_{n_{j\neq i}=0,1;
n_i=1}\delta_{n_1n_2n_3}$, where
$\delta_{n_1n_2n_3}=\sqrt{\det[w(\Gamma_{n_1n_2n_3})]}/{\rm
Pf}[w(\Gamma_{n_1n_2n_3})]=\pm 1$. Here the unitary matrix $w$ is
defined as $w_{ij}({\bf k})=\langle u_i(-{\bf k})|\Theta|u_j({\bf
k}\rangle$ with $\Theta$ being the time reversal operator and
$|u_j({\bf k})\rangle $ being the Bloch wave functions for occupied
bands. Fu and Kane have found  a simple method to identify the $Z_2$
invariants for the system with the presence of  inversion
symmetry\cite{Fu2}. In this case, $\delta_{n_1n_2n_3}$ can be
calculated by
$\delta_{n_1n_2n_3}=\prod_{m=1}^N\xi_{2m}(\Gamma_{n_1n_2n_3})$,
where $N$ is the number of occupied bands and
$\xi_{2m}(\Gamma_{n_1n_2n_3})=\pm 1$ is the parity eigenvalue of the
$2m$th occupied band at $\Gamma_{n_1n_2n_3}$. Our model is inversion
symmetric so we will adopt this method to evaluate the $Z_2$
invariants $\nu_i (i=0,1,2,3)$. We select the center of an
octahedron in the lattice as the center of inversion, then the
parity operator acts as ${\cal P}[\psi_1({\bf r}), \psi_2({\bf r}),
\psi_3({\bf r}),\psi_4({\bf r}), \psi_5({\bf r}), \psi_6({\bf r})]^T
=[\psi_6(-{\bf r}), \psi_4(-{\bf r}), \psi_5(-{\bf r}), \psi_2(-{\bf
r}), \psi_3(-{\bf r}), \psi_1(-{\bf r})]^T $, where  $[\psi_1({\bf
r}), \psi_2({\bf r}), \psi_3({\bf r}),\psi_4({\bf r}), \psi_5({\bf
r}), \psi_6({\bf r})]^T$ is the six-component wave function. Taking
Fourier transformation, we can write the six-component wave function
as $[\psi_1({\bf r}), \psi_2({\bf r}), \psi_3({\bf r}),\psi_4({\bf
r}), \psi_5({\bf r}), \psi_6({\bf r})]=\sum_{\bf k}[\phi_1({\bf k}),
\phi_2({\bf k}), \phi_3({\bf k}), \phi_4({\bf k}), \phi_5({\bf k}),
\phi_6({\bf k})]e^{i{\bf k}\cdot {\bf r}}$
 and the  parity operator as ${\cal P}=\sum_{\bf k} e^{i{\bf k}\cdot {\bf r}}{\cal P}_{\bf k} e^{-i{\bf
k}\cdot{\bf r}}$. Then, in momentum space, we obtain the equation
${\cal P}_{\bf k} [\phi_1({\bf k}), \phi_2({\bf k}), \phi_3({\bf
k}), \phi_4({\bf k}), \phi_5({\bf k}), \phi_6({\bf k})]^T
=[\phi_6(-{\bf k}), \phi_4(-{\bf k}), \phi_5(-{\bf k}), \phi_2(-{\bf
k}), \phi_3(-{\bf k}), \phi_1(-{\bf k})]^T$. Considering the degree
of spin,  we can express the parity operator at the time reversal
invariant momenta $\Gamma_{n_1n_2n_3}$ as follows,
\begin{eqnarray}
{\cal P}_{\Gamma_{n_1n_2n_3}}=\left(\matrix{1&0\cr
0&1}\right)\otimes\left(\matrix{0&0&0&0&0&1\cr 0 &0&0&1&0&0\cr
0&0&0&0&1&0\cr 0&1&0&0&0&0\cr 0&0&1&0&0&0\cr 1&0&0&0&0&0}\right)
\end{eqnarray}
where the $4\times 4$ matrix is the unit matrix in spin space.

We diagonalize the total single-particle Hamiltonian ${\cal H}_{\bf
k}$ and calculate the $Z_2$ topological invariants for different
filling fractions. We find that non-trivial topological insulators
exist for $1/6, 1/2$ and $2/3$ filling while only metal  phase
occurs for $1/3$ and $5/6$ filling. Thus, we will focus on and
discuss  the cases with $1/6, 1/2$ and $2/3$ filling fractions in
the following part of the paper. We identify phases for different
parameters $t_1$ and $\lambda_{\rm SO}$ with  $1/6, 1/2$ and $2/3$
filling fractions and draw phase diagrams as shown in
Fig.\ref{fig2}. Figs.\ref{fig2}(a), \ref{fig2}(b) and \ref{fig2}(c)
show the phase diagrams for $1/6, 1/2$ and $2/3$ filling,
respectively. For $1/6$ and $2/3$ filling, there are $(1; 111)$ and
$(1; 000)$ STI phases, $(0; 111)$ WTI phase as well as trivial band
insulator and metal phases.  For $1/2$ filling, there are $(1; 111)$
and $(1; 000)$ STI phases, trivial band insulator and metal phases
except   $(0; 111)$ WTI phase.

\begin{figure}[ht]
\includegraphics[width=1.1\columnwidth]{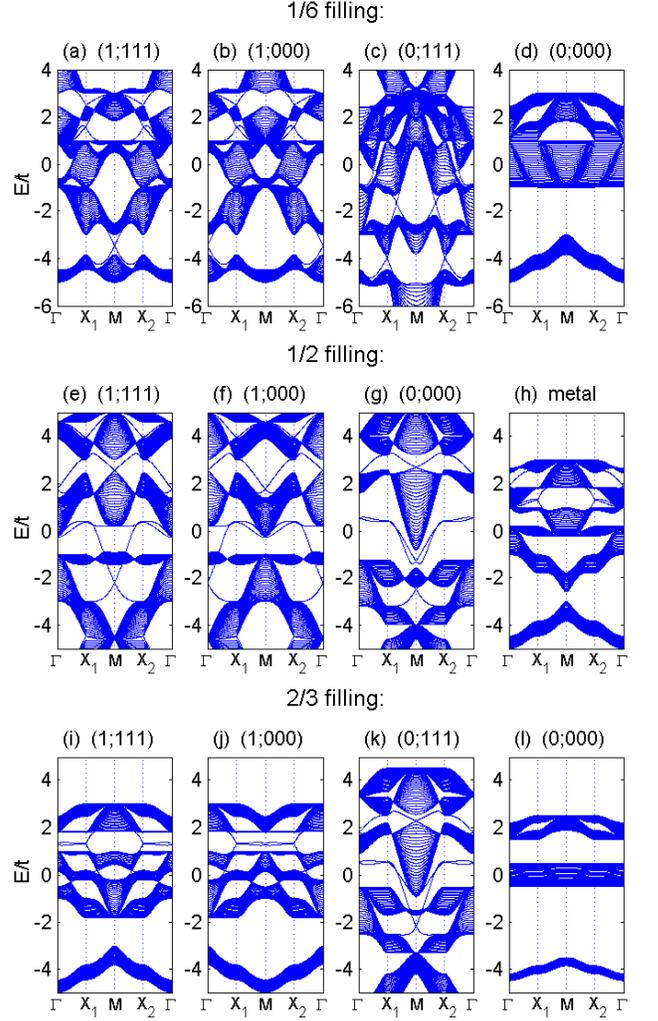}
\caption{(Color online). Band structures of a slab with two $001$
surfaces for various parameters $t_1$ and $\lambda_{\rm SO}$. Here,
the horizontal axis represents the wave vectors along the path in
the surface Brillouin zone indicated by the red lines in
Fig.\ref{fig1}(d). (a) $t_1=t, \lambda_{\rm SO}=0.5t$, (b) $t_1=-t,
\lambda_{\rm SO}=-0.5t$, (c) $t_1=t, \lambda_{\rm SO}=t$, (d)
$t_1=t, \lambda_{\rm SO}=0$, (e) $t_1=3t, \lambda_{\rm SO}=-0.4t$,
(f) $t_1=-3t, \lambda_{\rm SO}=0.4t$, (g) $t_1=3.2t, \lambda_{\rm
SO}=-0.2t$, (h) $t_1=t, \lambda_{\rm SO}=0.2t$, (i) $t_1=t,
\lambda_{\rm SO}=-0.2t$, (j) $t_1=-t, \lambda_{\rm SO}=0.2t$, (k)
$t_1=2.5t, \lambda_{\rm SO}=-0.2t$, and (l) $t_1=0.5t, \lambda_{\rm
SO}=0$. } \label{fig4}
\end{figure}

To clearly manifest the bulk band structure of different phases for
various filling factions, we calculate the bulk energy bands for
several cases with different parameters $t_1$ and $\lambda_{\rm
SO}$, which are shown in Fig.\ref{fig3}. In order to investigate the
characteristics of surface states for various phases, we evaluate
the energy bands in a slab geometry with two $001$ surfaces. The
Brillouin zone of the slab is shown in Fig.\ref{fig1}(d). The energy
bands are present along lines that connect the four surface TRIM as
shown in Fig.\ref{fig4}. With the assistance of the bulk energy
bands shown in Fig.\ref{fig3} and the two-dimensional energy bands
for a slab shown in Fig.\ref{fig4}, we will sequentially analyze
various phases, identify three-dimensional topological insulators,
and discuss their characteristics for $1/6, 1/2$ and $2/3$ filling.

\subsection{$1/6$ filling}

Fig.\ref{fig2}(a) shows the phase diagram of the
octahedron-decorated cubic lattice for $1/6$ filling. In this case,
the $(1;111)$ and $(1;000)$ STI phases are discovered. The
non-trivial STI phases have a gap between the first and second bands
as shown in Fig.\ref{fig3}(a) and \ref{fig3}(b) corresponding to
$(1;111)$ and $(1;000)$ STI phases, respectively. We note that  for
$1/6$ filling there is only one Dirac point on TRIM as shown in
Fig.\ref{fig4}(a) and (b), that is, only a pair of robust
spin-filtered states exists. We also find a $(0; 111)$ WTI phase for
$1/6$ filling e.g., as shown in Fig.\ref{fig3}(c). Fig.\ref{fig4}(c)
shows the surface states for a $(0; 111)$ WTI phase that has two
Dirac points between the first and second bands on TRIM. We note
that trivial band insulators occur for smaller $t_1$ and smaller
$\lambda_{\rm SO}$ parameters, which is easily understood for when
$t_1$ and $\lambda_{\rm SO}$ approaches to zero the lattice becomes
separated octahedral clusters. For a trivial band insulator  there
is a gap between the first and second bands as shown in
Fig.\ref{fig3}(d), but there are not surface states as shown in
Fig.\ref{fig4}(d). For a metal phase, the gap vanishes.

\subsection{$1/2$ filling}

Fig.\ref{fig2}(b) shows the phase diagram of  the
octahedron-decorated cubic lattice for $1/2$ filling. For $1/2$
filling, $(1;111)$ and $(1;000)$ STI phases, trivial band
insulators, and metal phases  occur, but WTI phases are not found.
Figs.\ref{fig3}(e) and \ref{fig3}(f) show the band structure for
$(1;111)$ and $(0;111)$ STI phases, respectively. We can find from
these diagrams that a gap opens between the third and fourth bands.
For STI phases, there only one Dirac point on TRIM as shown in
Fig.\ref{fig4}(e) and (f). For trivial band insulators, there is
also a gap between the third and fourth bands as shown in
Fig.\ref{fig3}(g), but even number of Dirac points exist on TRIM as
shown in Fig.\ref{fig4}(g). For smaller $t_1$ and smaller
$\lambda_{\rm SO}$, a metal phase occurs except a special point
$t_1=0$ and $\lambda_{\rm SO}=0$. For $t_1=0$ and $\lambda_{\rm
SO}=0$, the second, third and fourth bands are degenerate and become
a flat band, which means that electrons are localized. In other
works, the system with $t_1$ and $\lambda_{\rm SO}$ for $1/2$
filling is a trivial band insulator. However, a tiny change from
$t_1=0$ and $\lambda_{\rm SO}=0$ for parameters $t_1$ and
$\lambda_{\rm SO}$ makes the flat band become three dispersive bands
that are crossover each other, then the lattice with three bands
occupied becomes a metal.

\subsection{$2/3$ filling}

Fig.\ref{fig2}(c) shows the phase diagram  of the
octahedron-decorated cubic lattice for $2/3$ filling. We note that,
similar to $1/6$ filling,  $(1;111)$ and $(1;000)$ STI phases,
  $(0;111)$ WTI phase, trivial band insulator, and metal phase
occur in different ranges of parameters $t_1$ and $\lambda_{\rm
SO}$. Fig.\ref{fig3}(i) shows the band structure for $t_1=t,
\lambda_{\rm SO}=-0.2t$ at which a $(1;111)$ STI phase occurs. We
can find that a gap opens between the fourth and fifth bands as
shown in Fig.\ref{fig3}(i). There is an odd number of surface states
which traverse the gap as shown in Fig.\ref{fig4}(i). For the
$(1;000)$ STI phase, the similar characteristics are exemplified in
Figs.\ref{fig3}(j) and \ref{fig4}(j). The $(0;111)$ WTI phase is
found as well. Fig.\ref{fig3}(k) and Fig.\ref{fig4}(k) show the
$(0;111)$ WTI phase has a gap between the fourth and fifth bands and
an even number of surface states traversing the gap.
 For smaller $t_1$ and smaller $\lambda_{\rm
SO}$, the system for $2/3$ filling is a trivial band insulator,
which is feathered by a gap between the fourth and fifth bands
combined with an even number of surface states traversing the gap as
shown in Fig.\ref{fig3}(l) and Fig.\ref{fig4}(l), respectively.

\section{Conclusion}

In summary, we have shown that the octahedron-decorated cubic
lattice with spin-orbit coupling supports three-dimensional
topological insulators at  $1/6, 1/2$ and $2/3$ filling fractions.
For $1/6$ and $2/3$ filling,  $(1;111)$ and $(1;000)$ STI phases,
$(0;111)$ WTI phase, trivial band insulator, and metal phase are
found, while for $1/2$ filling,   $(1;111)$ and $(1;000)$ STI
phases, trivial band insulator, and metal phase occur except
$(0;111)$ WTI phase. We have calculated the band structure and
surface band structure for the tight-binding model of the
octahedron-decorated cubic lattice with spin-orbit coupling and
evaluated the $Z_2$ topological invariants. We have analyzed and
discussed the characters of the band structures and the surface
states of different phases. Although the octahedron-decorated cubic
lattice we considered is a toy model, our study points out an
alternative path to search for real topological materials. On the
other hand, it might as well be built from optical lattices due to
their diversity and controllability.

\begin{acknowledgments}
This work was supported  by the National Natural Science Foundation
of China under Grant No. 11004028 and the Science and Technology
Foundation of Southeast University under Grant No. KJ2010417
\end{acknowledgments}

\end{document}